\documentclass{aa}  

\usepackage{graphicx}
\usepackage{txfonts}
\usepackage{color}
\usepackage[]{hyperref}

\newcommand{\xmm}{\emph{XMM--Newton}~}
\newcommand{\pc}{\phantom{0}}

\begin{document}

   \title{Disc wind or disc line? The extraordinary Fe-K feature of Mrk 1513}


   \author{R. Middei 
          \inst{1,2}\fnmsep\thanks{riccardo.middei@ssdc.asi.it}
          \and
           E. Nardini \inst{3}
           \and G. A. Matzeu  \inst{4}
           \and S. Bianchi \inst{5}
           \and V. Braito \inst{6,7,8}
           \and M. Perri \inst{1,2}
           \and S. Puccetti \inst{2}
          }

   \institute{ 
   INAF Osservatorio Astronomico di Roma, Via Frascati 33, 00078 Monte Porzio Catone (RM), Italy         \and 
   Space Science Data Center, Agenzia Spaziale Italiana, Via del Politecnico snc, 00133 Roma, Italy
         \and INAF -- Osservatorio Astrofisico di Arcetri, Largo Enrico Fermi 5, I-50125 Firenze, Italy
         \and Quasar Science Resources SL for ESA, European Space Astronomy Centre (ESAC), Science Operations Department, 28692, Villanueva de la Ca\~{n}ada, Madrid, Spain
         \and Dipartimento di Matematica e Fisica, Universit\`a degli Studi Roma Tre, Via della Vasca Navale 84, 00146 Roma, Italy
         \and Department of Physics, Institute for Astrophysics and Computational Sciences, The Catholic University of America, Washington, DC 20064, USA
         \and INAF, Osservatorio Astronomico di Brera, Via Bianchi 46, I-23807 Merate (LC), Italy
        \and Dipartimento di Fisica, Università di Trento, Via Sommarive 14, I-38123 Trento, Italy}


   \date{Received mm/dd/yyyy; accepted mm/dd/yyyy}

 
\abstract  
   {We discuss the origin of a very unusual spectral structure observed in the Fe-K band of the Seyfert galaxy Mrk 1513, a local ($z$\,=\,0.063) active galactic nucleus (AGN) that is efficiently accreting matter onto its central supermassive black hole ($L_{\rm bol}$/$L_{\rm Edd}$\,$\sim$\,0.5). We consider the highest quality X-ray observation of this source available to date, performed in 2003 by {\it XMM--Newton}. The hard X-ray spectrum is characterised by a remarkable spectral drop at $\sim$7 keV, which can be interpreted as either the onset of a broad absorption trough or the blue wing of a relativistic emission line. Overall, this complex feature is significant at >\,5$\sigma$, and it is qualitatively reminiscent of a P-Cygni profile. 
   A serendipitous spectrum of lower quality taken by {\it XMM--Newton} in 2015 qualitatively confirms the presence of similar Fe-K structures.
   Although it is not possible to distinguish between the two physical scenarios on sheer statistical grounds with the current data, several considerations lend weight to the possibility that Mrk\,1513 is actually hosting a persistent outflow at accretion-disc scales, thus adding to the handful of known AGN in which a wide-angle X-ray wind has been identified so far.}
   



    \keywords{galaxies: active -- galaxies: Seyfert -- X-rays: galaxies -- X-rays: individual: Mrk\,1513}

   \maketitle
%


\section{Introduction}

It is generally accepted that accreting supermassive black holes (SMBHs) in active galactic nuclei (AGN) play a key role in shaping the evolution of their host galaxies \citep[][]{Kormendy2013}. Feedback in the form of powerful mass outflows is thought to energetically connect these two spatially 
independent regions \citep[][]{Laha2021N}, as 
the propagation through the interstellar medium of the so-called ultra-fast outflows generates forward and backward shocks \citep[e.g.,][]{Faucher2012} that can enhance or quench the star formation in the host galaxy \citep[e.g.,][]{Sturm2011,Zubovas2013,Zubovas2014,Cresci2015}.
In 
AGN accreting with high efficiency, powerful winds 
with high column density ($N_{\rm H}$\,>\,10$^{23}$ cm$^{-2}$), ionisation ($\log \xi$\,>\,4)\footnote{The ionisation parameter, defined as $\xi=L_{\rm ion}/nr^2$ (erg cm s$^{-1}$), depends on the ratio between the ionising flux received by the gas and its density. For simplicity, its units are omitted throughout the text.}, 
and velocity ($-0.4c$\,$\la$\,$v_{\rm out}$\,$\la$\,$-0.1c$) are expected from both theoretical arguments \citep[e.g.,][]{King2003} and simulations 
\citep[e.g.,][]{Proga2004}. From an observational perspective, these outflows imprint X-ray absorption features at blueshifted energies, usually in the Fe-K band (i.e., Fe\,\textsc{xxv} and Fe\,\textsc{xxvi}). 

Despite plenty of studies reporting on ultra-fast X-ray winds in nearby Seyferts and quasars, several open questions remain, such as the launching and driving mechanisms, and the geometry and duty cycle of these outflows. In particular, the detection rate of Fe-K absorption in relatively large samples of local AGN 
has been often used as an indirect measurement of the covering factor of disc winds, and these ensemble-based estimates in the range 0.4--0.6 \citep[][see also \citealt{Igo2020}]{Tombesi2010,Gofford2013} suggest the wind to cover about half of the sky as seen by the X-ray continuum source. This is a key issue, as the degree of collimation of the ejected matter, combined with the sheer kinetic power, is a critical parameter to achieve an effective feedback on large scales. 
In the case of outflows with such a wide opening angle, a typical P-Cygni spectral feature would be expected \citep[e.g.,][]{Castor1979}. 
However, this 
distinctive emission--absorption profile is exceptionally rare in the X-ray spectra of AGN, so that a clear detection of a P-Cygni feature, or claims thereof, 
only exist for a few objects, including PG\,1211+143 \citep{pounds2009}, PDS\,456 \citep{Nardini2015}, and I\,Zw\,1 \citep[][]{Reeves2019}. Tentative identifications of this profile were also reported in some high-redshift objects \citep[e.g.,][]{Vignali2015,Dadina2016}.

The picture is further complicated by the possible spectral degeneracy between outflow and reflection signatures. Indeed, even in the presence of blueshifted Fe-K absorption, any emission counterpart can be also associated with the reflection of the primary X-ray radiation off the surface of the accretion disc. This makes it very difficult to disentangle the relative disc and wind contributions and so extract any sensible physical information from the observed line profile \citep{Parker2022}. It is worth noting that such a degeneracy works in both directions, as the extreme physical parameters implied by the strongest disc lines, among which a largely supersolar Fe abundance, can be significantly softened by allowing for the presence of a disc wind \citep[e.g.,][]{Hagino2016}.

A high Eddington rate is believed to imply favourable conditions for a fast X-ray wind to develop 
\citep[e.g.,][]{Reynolds2012}, as the optical depth to Thomson scattering is high enough for continuum radiation pressure to sustain an outflow without requiring any substantial force multiplier associated to line absorption, which is negligible at high ionisation (but see \citealt{Hagino2015}). The importance of accretion rate and radiation driving is also revealed by the correlation between outflow velocity and X-ray luminosity found by \citet[][]{Matzeu2017} in PDS\,456, while in NGC\,2992 the wind signatures only emerge when the Eddington rate exceeds $\sim$2\% \citep{Marinucci2018}. We refer to \citet{Giustini2019} for a more general, holistic picture of the various wind manifestations as a function of the accretion rate. We note, however, that 
large uncertainties affect the estimate of the latter parameter, and it is therefore desirable to rely on BH masses derived via reverberation mapping techniques. 

The ideal targets to gain deeper insights into the physics of accretion-disc winds are therefore AGN for which a sensible determination of BH mass and accretion rate is available, and possibly showing evidence of outflows also in other gas phases of lower ionisation. In this context, we revisit two archival \textit{XMM--Newton} observations of the 
Seyfert 1 galaxy Mrk\,1513 (RA\,=\,21h\,32m\,27.8s, Dec\,=\,+10$^{\circ}$\,08$'$\,19$''$; also known as PG\,2130+099 or UGC\,11763), at redshift $z=0.063$ \citep[][]{Springob2005}.
For this AGN, \citet{Grier2017A} inferred a BH mass of $\log M_{\rm BH}$\,=\,$6.92_{-0.23}^{+0.24}\,M_\sun$ by modelling the reverberation mapping data of the broad H$\beta$ component (see also \citealt{Hu2020}). By applying the bolometric correction of \citet{Duras2020} to the hard X-ray luminosity of Mrk\,1513, $L_{\rm 2-10\,keV}$\,=\,$3.5\times10^{43}$ erg s$^{-1}$, we derive an accretion rate of $L_{\rm bol}/L_{\rm Edd}\sim0.5$. A nearly identical value is obtained by considering the 5100-\AA\ luminosity with the standard bolometric correction factor of 9 (after \citealt{Richards2006}). Mrk\,1513 can then be safely assumed to belong to the high-efficiency accretion regime.


Contrary to the extensive optical monitoring, only a few studies focussed on this AGN in the X-rays. \citet{Cardaci2009} reported on 
the presence of two warm-absorption components with moderate ionisation and column densities of a few $\times$10$^{21}$ cm$^{-2}$, mostly responsible for absorption by the Fe M-shell unresolved transition array \citep[e.g.,][]{Sako2001}. Both components appear to be inflowing at 500$\pm$300 km s$^{-1}$, yet the same authors caution against over-interpreting this result as the low signal-to-noise degraded the nominal resolution of the Reflection Grating Spectrometers (RGS, \citealt{denHerder2001}) onboard \textit{XMM--Newton}. Interestingly, the \textit{Hubble Space Telescope} ({\it HST}) Cosmic Origins Spectrograph data of Mrk\,1513 show an intrinsic narrow absorption line (NAL) system with an outflow velocity $v_{\rm out}$\,=\,$-1521\pm20$ km s$^{-1}$, seen in Ly$\alpha$ and high-ionisation species like N\,\textsc{v}, Si\,\textsc{iv}, and C\,\textsc{iv} \citep[][and references therein]{Tofany2014}, implying that a stream of hot gas is ejected at high velocity from the nuclear regions of this AGN. Here we discuss the intriguing possibility that Mrk\,1513 actually hosts even an ultra-fast disc wind. This letter is organised as follows: Section 2 concerns the \xmm observations and data reduction, Section 3 describes the results of the spectral analysis, while our conclusions are drawn in Section 4.

%
\section{\textit{XMM--Newton} observation and data reduction}
Mrk\,1513 was first observed with the EPIC pn and MOS cameras \citep[][respectively]{Struder2001,Turner2001} onboard \xmm \citep[][]{Jansen2001} on 2003 May 17 for 37.5 ks (ObsID 0150470701). This remains the highest quality X-ray observation available for this AGN to date. Indeed, the source was also serendipitously observed on 2015 May 1 for 30 ks, yet in the latter data set (ObsID 0744370201) Mrk\,1513 lies about 5$\arcmin$ off-axis and falls in the gap between two chips of the pn CCD. As this unfortunate position significantly affects the usefulness of the 2015 observation for the present analysis (Table~\ref{tabobs}), we will mainly focus on the 2003 spectrum. However, the main features discerned in the Fe-K band are remarkably similar in the two epochs, at least qualitatively (see Section \ref{xmmobs}). 

The two \xmm exposures were reprocessed using the standard pipeline within the Science Analysis System (\textsc{sas}, version 20.0.0) and the latest calibration files (March 2023). 
Except for the 2015 MOS data, all the other exposures are somewhat spoiled by some background flares. These periods were filtered via an iterative process aimed at maximising the S/N ratio in the 4--9 keV energy band \citep[e.g.,][]{Piconcelli2004,Nardini2019}. We extracted the source spectra from circular regions with radius of 30$\arcsec$ (2003) or 35$\arcsec$ (2015) for the pn, and 25$\arcsec$ for both MOS detectors. The background regions were selected in order to avoid nearby sources and possible instrumental features (due to, e.g., Cu-K fluorescence), with a circular shape\footnote{The background region is instead a 20$\arcsec$\,$\times$\,100$\arcsec$ rectangular strip for the 2003 MOS1 data, which were acquired in Small Window mode.} and radii between 60$\arcsec$ and 75$\arcsec$. The basic details of each observation are listed in Table~\ref{tabobs}. Response files were produced with the \textsc{sas} tools \texttt{rmfgen} and \texttt{arfgen}. Finally, the obtained spectra were binned in order to have at least 5 counts per bin. 
The \textsc{sas} routine \texttt{epicspeccombine} was used to sum the EPIC spectra together. The output files, however, are only employed for plotting purposes, unless stated otherwise.

\begin{figure}
   \centering
   \includegraphics[width=\columnwidth]{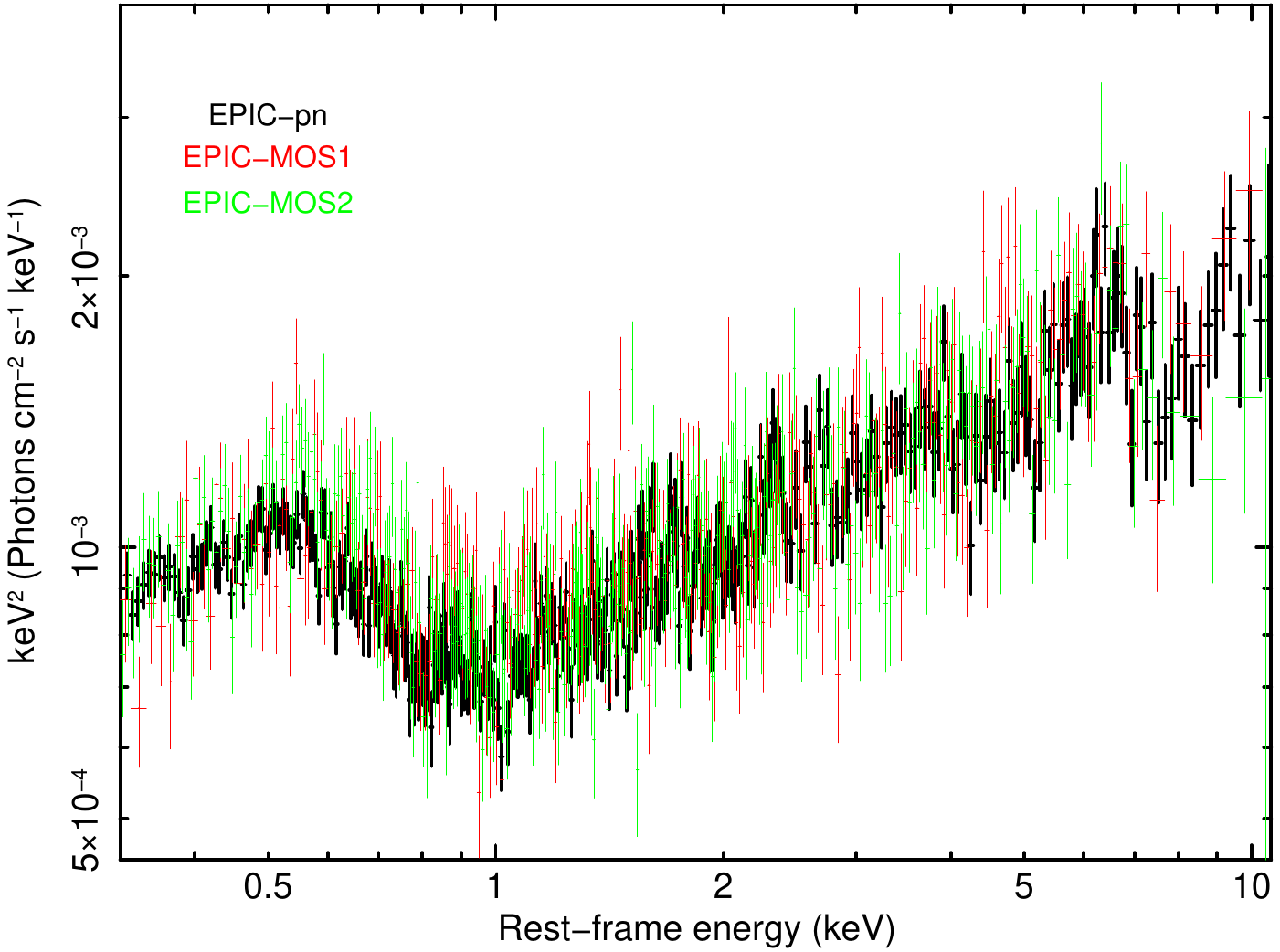}
      \caption{EPIC-pn (black), MOS1 (red), and MOS2 (green) spectra from the 2003 \xmm observation of Mrk\,1513, unfolded against a power law with $\Gamma$\,=\,2. Besides the soft excess component below 1 keV, all the spectra are characterized by a sharp drop around 7 keV. The emission deficit up to $\sim$9 keV is clearly noticeable in the pn data, but also confirmed by the two MOS detectors despite their rapidly declining effective area at these energies.}
         \label{unfolded}
   \end{figure}

In Fig.~\ref{unfolded}, we show the EPIC spectra of Mrk\,1513 derived for the 2003 \xmm observation, unfolded against a power law with $\Gamma=2$ and normalisation equal to unity. A sharp spectral drop is clearly visible at $\sim$7 keV. A similar feature is present during the 2015 observation, despite the lower hard X-ray statistics and a different broadband spectral shape (see below).

\begin{table*}
\centering
\caption{Log of the 2003 (ObsID 0150470701) and 2015 (ObsID 0744370201) \xmm observations of Mrk\,1513.}
\label{tabobs}
\setlength{\tabcolsep}{0pt}
\begin{tabular}{l@{\hspace{35pt}}ccc@{\hspace{30pt}}ccc@{\hspace{30pt}}c}
\hline
Date & & Exposure$^a$ (s) & & & Counts$^b$ (5--10 keV) & & Flux$^c$ (0.3--10 keV) \\
 & pn & MOS1 & MOS2 & pn & MOS1 & MOS2 & 10$^{-11}$\,erg\,s$^{-1}$\,cm$^{-2}$ \\
\hline \noalign{\smallskip}
2003 May 16 & 24267 & 29764 & 36920 & 2478$\pm$52 & 685$\pm$28 & 857$\pm$30 & 0.656$\pm$0.004 \\
2015 May 01 & 25599 & 31236 & 31220 & \pc688$\pm$28 & 622$\pm$25 & 657$\pm$26 & 1.086$\pm$0.007 \\
\hline
\end{tabular}
\flushleft
\textit{Notes.} $^a$Net exposure after dead-time correction and filtering of high-background periods. $^b$Net counts in the hard X-ray band; note the major shortcoming with the 2015 pn data. $^c$Observed flux (not corrected for Galactic absorption) computed through a phenomenological spectral model. The difference is mostly due to the intensity of the soft X-ray emission below $\sim$3 keV.\label{log}
\end{table*}

\section{Spectral analysis}
\label{span}

The spectral analysis presented here was performed with the fitting software package \textsc{xspec} \citep[][]{Arnaud1996}. 
In all the fits the C-statistic \citep{Cash1979} was adopted and Galactic column density is always included via the \texttt{tbabs} model \citep[$N_{\rm H}=3.7\times 10^{20}$ cm$^{-2}$,][]{HI4PI2016}. 
We also used a constant component free to vary to account for the cross-normalisation between the different EPIC detectors, which is within 5\%.

\citet{Cardaci2009} analysed the 2003 \xmm observation of Mrk\,1513 and found 
the RGS spectra to reveal the presence of warm-absorption features. Two mildly ionised components with a marginal redshift (hence inflowing) were statistically preferred over a single, blueshifted one (at $v_{\rm out} \simeq -300$ km s$^{-1}$). 
As our aim is to investigate the nature of the spectral structures in the Fe-K band, and the velocity of these warm absorbers cannot be constrained at CCD resolution, in our analysis we assumed that both components are at rest in the local frame. Then, both the column density and the ionisation parameter $\xi$ were calculated during the fitting procedure. In any case, the Fe-K band is insensitive to the best-fit properties of the warm absorbers. 



The sharp drop observed at 7 keV in the \xmm spectra of Mrk\,1513 can be alternatively interpreted as due to either the blue horn/wing of a relativistic disc line, or an Fe-K absorption edge \citep[e.g.,][]{Boller2002}. 
To better visualize the shape of this emission--absorption structure, we fitted the 0.3--10 keV spectrum with the following \textsc{xspec} model: \texttt{tbabs\,$\times$\,xabs\,$\times$\,xabs\,$\times$\,(bbody\,+\,powerlaw)}, where the two \texttt{xabs} components\footnote{This is a table implemented for \textsc{xspec} by \citet{Parker2019}, derived from the homonym photoionised absorption model \citep[see][]{Steenbrugge2003} in \textsc{spex} \citep[][]{Kaastra1996}. The table is available at the webpage \url{https://www.michaelparker.space/xspec-models}.} account for the warm absorber, and a phenomenological blackbody and a power law are used to reproduce, respectively, the soft excess and the hard nuclear continuum (compare with Fig.~\ref{unfolded}). All the parameters of the blackbody and power-law components were left free to vary. This simple model returns a statistics of C-stat\,=\,3222 for 3131 degrees of freedom (d.o.f.) and, unsurprisingly, it leaves prominent residuals above 6 keV. We then carried out a blind line scan by adding to the underlying continuum model a Gaussian profile with fixed width of 150 eV (i.e., compatible with the CCD resolution). The line energy and normalisation were allowed to vary with 50 steps from 5.5 to 9.5 keV and from $-$1.5 to 1.5\,$\times$\,10$^{-5}$ photons cm$^{-2}$ s$^{-1}$, respectively. 

\begin{figure}
   \centering
\includegraphics[width=\columnwidth]{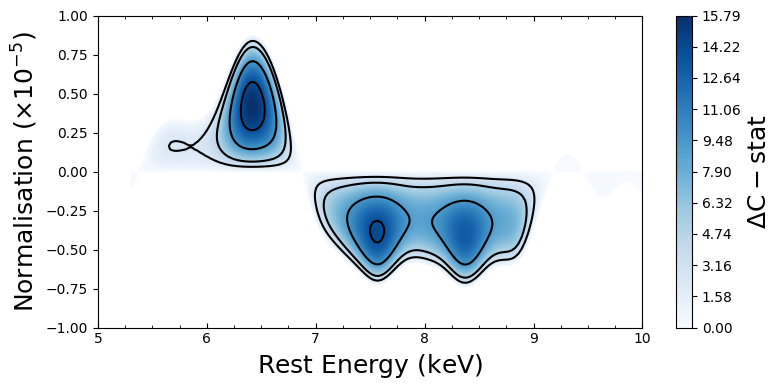}
      \caption{Results of the blind scan for the line normalisation and energy based on a phenomenological continuum model (see the text for more details). 
      The contours correspond to the 68\%, 90\%, 99\%, and 99.9\% confidence levels. The proximity in energy of positive and negative residuals is strongly reminiscent of a P-Cygni profile.}
         \label{linescan}
\end{figure}

The outcome of this procedure is shown in Fig.~\ref{linescan}, which reveals two broad features, one in emission and the other one in absorption, whose combination is strongly reminiscent of a P-Cygni profile. However, if the continuum is more complex than assumed in our baseline model, or moderately bent at higher energies, the same residuals could also be explained by a single relativistic emission line. We thus explore both the wide-angle wind and the disc reflection scenarios in more depth below.

\subsection{An ultra-fast disc wind?}
\label{ufo}
According to the line scan, the 7-keV spectral drop is apparently due to the presence of a broad absorption trough immediately blueward of a prominent emission feature, whose peak is compatible with fluorescence from iron in a low-ionisation state. The trough, which is tempting to associate with an accretion-disc wind, possibly has a complex structure with two dips astride 8 keV, although the higher energy one is only marginally significant. Based on Fig. \ref{linescan}, we added to the baseline continuum two Gaussian components with, respectively, positive and negative amplitude, which yield a statistical improvement of $\Delta$C-stat/$\Delta$d.o.f.\,$\approx$\,$-$45/$-$6. The probability of chance improvement is $<$\,10$^{-7}$, and the aggregate feature is significant at $>$\,5$\sigma$. Yet, the centroids and widths of both lines are hard to constrain simultaneously in a reasonable way, as they are all highly degenerate with each other; in fact, the two profiles substantially overlap in order to reproduce the steepness of the flux drop as well as the mild asymmetry of the emission and absorption tails. 

Scattering by outflowing matter that covers a large fraction of the solid angle can give rise to intense Fe-K emission at lower energies than the absorption trough \cite[e.g.,][]{Nardini2015}. 
As the Gaussian-line fit is poorly informative, we thus tested a self-consistent P-Cygni profile adopting the \texttt{pcygx} model, customized for \textsc{xspec} by \cite{Done2007} after \cite[][]{Lamers1987}. This model mimics the simplified scenario of an expanding, spherically symmetric wind with a velocity field of the form 
$w$\,=\,$w_0$\,+\,(1\,$-$\,$w_0$)(1\,$-$\,$1/x)^{\gamma}$, where
$w$\,=\,$v/v_{\inf}$ is the velocity of the wind normalised to the terminal velocity, $w_0$ is the dimensionless velocity of the wind photosphere, $x$ represents the ratio between the radial distance $r$ and the radius of the photosphere, and $\gamma$ sets the scaling relation between distance and velocity of the wind. 
Both $w_0$ 
and $\gamma$ have negligible impact on the fits, hence we fixed their values to 0.01 and 1, respectively. Under these assumptions, the line optical depth can be written as $\tau(w)$\,$\propto$\,$\tau_{\rm tot}w^{\alpha_1}$(1$-$\,$w)^{\alpha_2}$, where $\alpha_1$ and $\alpha_2$ regulate the P-Cygni profile smoothness (see \citealt{Ramirez2005}). 
We also assumed for simplicity that $\alpha_1$\,=\,$\alpha_2$\,=\,0, and considered as free parameters the rest-frame energy of the line $E_0$, the terminal velocity of the wind $v_{\inf}$, and the total optical depth of the gas $\tau_{\rm tot}$. 

This model provides a satisfactory representation of the Fe-K spectrum of Mrk\,1513, with a statistics comparable to the double-Gaussian model but three more degrees of freedom, i.e., C-stat/d.o.f.\,=\,3176/3128. The line rest energy is fully consistent with Fe\,\textsc{xxv} He$\alpha$ at $E_0$\,=\,6.72$^{+0.11}_{-0.14}$ keV, the gas optical depth is $\tau_{\rm tot}$\,$\sim$\,0.2, and the terminal velocity is $v_{\inf}$\,$\sim$\,0.34$c$. All the best-fit parameters are reported in Table~\ref{tabtext}, while best-fit model and residuals are shown in Fig.~\ref{finalebestfit} (top and middle panels).

\subsection{A reflection-dominated spectrum?}
\label{refl}

Fluorescence, scattering, and recombination are known to occur also from iron inside the disc, hence the emission feature at 6--7 keV is not an unambiguous signature of a wide-angle wind (see \citealt{Parker2022}). We therefore replaced the P-Cygni component with an emission line from a relativistic accretion disc, modelled through the \texttt{diskline} profile in \textsc{xspec} (e.g., \citealt{Fabian1989}). The free parameters of this component are the line energy and normalisation, and the disc inner radius and inclination. We eventually assumed a disc emissivity scaling with radius as (1\,$-$\,$\sqrt{6/R})/R^3$, where $R$ is in units of gravitational radii ($r_{\rm g}$\,=\,$GM_{\rm BH}/c^2)$, as this is preferred by the fit. 

\begin{figure}
\centering
\includegraphics[width=\columnwidth]{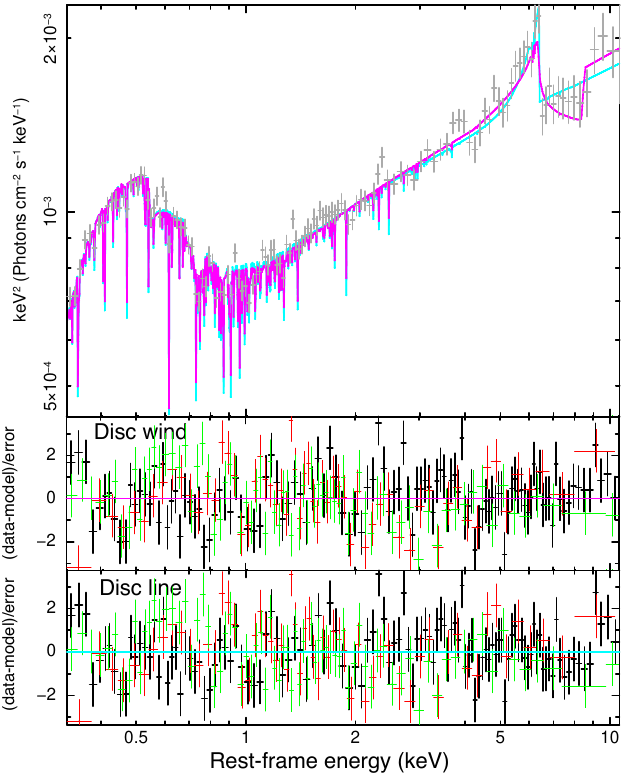}
 \caption{Top panel: \texttt{pcygx} (magenta) and \texttt{diskline} (cyan) best-fit models, superimposed for reference to the combined EPIC spectrum (rebinned for clarity). Note that the effects of the warm absorber on the Fe-K band are negligible in either case. 
 Middle panel: fit residuals with respect to \texttt{pcygx} model, computed as (data$-$model)/error for the individual EPIC spectra (colour-coded as in Fig. \ref{unfolded}). Bottom panel: same as above for the \texttt{diskline} model.}
\label{finalebestfit}
\end{figure}

As shown in Fig.~\ref{finalebestfit} (top and bottom panels), this model delivers an equally good fit to the \xmm spectrum of Mrk\,1513, with C-stat/d.o.f.\,=\,3173/3127. The best-fit parameters are listed in Table~\ref{tabtext}. The line rest energy of $E_0$\,=\,6.48$^{+0.06}_{-0.03}$ keV suggests that the gas in the disc is only moderately ionised. The disc is seen at intermediate inclination, $i$\,$\sim$\,40$\degr$, and extends down to the innermost stable circular orbit of a non-rotating black hole. Even without taking the line and disc parameters at face value, it is clear that also a relativistic line can effectively describe the complex 7-keV spectral structures observed in Mrk\,1513. We note, however, that any disc line is required to be very strong, with an equivalent width of EW\,=\,330\,$\pm$\,120 eV, whereas the average value for broad Fe-K lines is $\sim$100 eV \citep[e.g.,][]{Nandra2007,Patrick2012}. This entails strong light bending to efficiently illuminate the disc and suppress the fraction of direct continuum that reaches the observer at infinity \citep{MiniuttiFabian}. Hence, we would expect the broadband X-ray spectrum of Mrk\,1513 to be reflection-dominated, and we should be able, in principle, to account for both the Fe-K feature and the excess emission in the soft band within a relativistic reflection framework. 
To this aim, we performed several tests by taking advantage of the \textsc{relxill} suite \citep[e.g.,][]{Garcia2014,Dauser2016}, which allows one to reproduce the primary and reflected emission for various physical and geometrical configurations of the disc--corona system.

\indent

\begin{table}
\caption{Best-fit parameters derived for the alternative models adopted to reproduce the 7-keV spectral feature(s) of Mrk\,1513. Errors are given at the 90\% confidence level for the single parameter of interest.} \label{tabella}
\label{tabtext}
\begin{tabular}{lcc}
\hline
Model & P-Cygni & Disc line \\
\hline \noalign{\smallskip}
Galactic absorption (\texttt{tbabs}) & & \\
$N_{\rm H}$ (10$^{20}$\,cm$^{-2}$) & 3.7 & 3.7 \\[0.8ex]
Warm absorber (\texttt{xabs}) & & \\
$\log\xi_1$ & 1.5$\pm$0.2& 1.5$\pm$0.2\\
$N_{\rm H,1}$ (10$^{21}$\,cm$^{-2}$)& 1.0$\pm$0.4 & 1.2$\pm$0.4\\
$\log\xi_2$ & >\,2.9 & >\,2.9\\
$N_{\rm H,2}$ (10$^{21}$\,cm$^{-2}$)& 4.3$\pm$0.4 & 6.3$\pm$0.4\\[0.8ex]
Soft excess (\texttt{bbody}) & & \\
$T_{\rm bb}$ (eV) & 96$\pm$3 & 95$\pm$3\\
$\log F_{\rm 0.3-2~keV}$ &$-$11.90$\pm$0.01 &$-$11.88$\pm$0.02\\[0.8ex]
Hard continuum (\texttt{powerlaw}) & & \\
$\Gamma$ & 1.66$\pm$0.02 & 1.70$\pm$0.03\\
$\log F_{\rm 2-10~keV}$ & $-$11.42$\pm$0.01 &$-$11.43$\pm$0.01\\[0.8ex]
P-Cygni line (\texttt{pcygx}) & & \\
$E_{\rm 0}$ (keV) & 6.72$^{+0.11}_{-0.14}$ & -- \\
v$_{\inf}/c$ 
& 0.34$^{+0.06}_{-0.09}$ & -- \\
$\tau_{\rm tot}$ & 0.22$\pm$0.08 & -- \\[0.8ex]
Disc line (\texttt{diskline}) & & \\
$E_{\rm 0}$ (keV) & -- & 6.48$^{+0.06}_{-0.03}$ \\
$R_{\rm in}$ ($r_{\rm g}$) & -- & <\,21 \\
$i$ (degrees) & -- & 42$^{+12}_{-23}$ \\
$A$ (10$^{-5}$ cm$^{-2}$ s$^{-1}$) & -- & 1.3$\pm$0.3 \\
\hline
C-stat/d.o.f. & 3176/3128& 3173/3127\\
\hline

\end{tabular}
\end{table}

\subsection{Broadband relativistic reflection models}
\label{relx}
\begin{figure}
\centering
\includegraphics[width=\columnwidth]{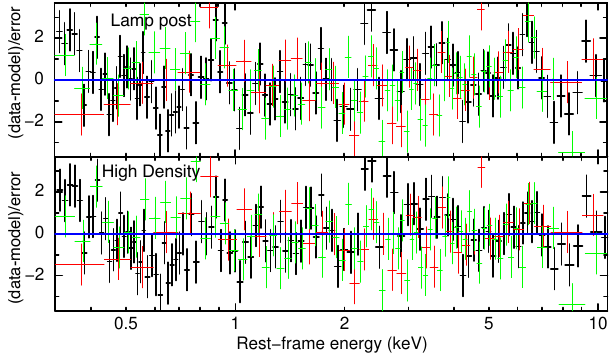}
 \caption{Fit residuals for the \texttt{relxilllp} (top) and \texttt{relxillD} (bottom) relativistic reflection models applied to the 2003 \xmm spectrum of Mrk\,1513. The colour code is the same as in Fig. \ref{unfolded}.}
\label{allratios}
\end{figure}

The \textsc{relxill} package provides a set of spectral models in which the relativistic reflection from an accretion disc illuminated by the primary X-ray is self-consistently calculated during the fitting procedure. Given the strength of the disc line required to explain the 7-keV drop, we tested whether any model in the \textsc{relxill} family allows us to account for the entire 0.3--10 keV spectrum of Mrk\,1513. We started by adopting the lamppost configuration for the primary X-ray source, as rendered through the following model expression: \texttt{tbabs\,$\times$\,xabs\,$\times$\,xabs\,$\times$\,relxilllp}, where the latter component is meant to incorporate both the direct and the reprocessed flux under the assumed geometry. We performed the fit computing the photon index $\Gamma$ of the primary continuum, the height of the corona above the disc ($h$, which is also set to control the reflection fraction), 
and the ionisation parameter $\xi_{\rm d}$, inclination $i$, and iron abundance $A_{\rm Fe}$ of the disc. The SMBH spin $a^*$ was also left as a free parameter under the hypothesis that the disc extends down to the innermost stable circular orbit. The continuum high-energy cutoff was frozen to the value of 300 keV, well beyond the \xmm operating energy range, and the two warm-absorption components were treated as described above. 
The \texttt{relxilllp} model yields a statistics of C-stat/d.o.f.\,=\,3245/3128, and it is moderately worse than the phenomenological continuum model. In particular, P-Cygni-like residuals remain in the Fe-K band (Fig.~\ref{allratios}, top panel). The inferred best-fit quantities are reported in Table~\ref{tabapp}.

This poor result can be ascribed to the lack of physical and geometrical flexibility of the \texttt{relxilllp} model. We then replaced it with the \texttt{relxillD} flavour, where the density of the disc is no more fixed at $\log\,(n/{\rm cm}^{-3})$\,=\,15 but becomes a free parameter. Moreover, the lamppost configuration is dismissed, allowing the disc emissivity to follow a broken-power-law dependence on radius (for simplicity, we assumed a trend $\propto$\,$r^{-q}$ with no break) and the reflection fraction $\mathcal{R}$ (see \citealt{Dauser2016}) to vary. With the emissivity exponent and reflection fraction as new free parameters in place of the coronal height, and the addition of the disc density, the fit improves down to C-stat/d.o.f.\,=\,3194/3126 (Table~\ref{tabapp}). The residuals in the Fe\,K band are somewhat attenuated but not completely offset (Fig.~\ref{allratios}, bottom panel). This suggests that, for the 7-keV feature to have a reflection origin, a different physical process must be responsible for the bulk of the soft X-ray emission in Mrk\,1513. Indeed, the fit is invariably driven by the soft excess, which calls for a degree of relativistic blurring that is far too large to simultaneously produce a sharp drop as the one seen around 7 keV in Mrk\,1513. We then argue that none of the broadband reflection-based models, while largely acceptable in terms of overall statistics, are able to account for the Fe-K spectral structure, leaving positive and negative residuals with a P-Cygni-like shape. 

\indent

\begin{table}
\centering
\caption{Best-fit parameters derived for the different relativistic reflection models applied to the 2003 EPIC spectra of Mrk\,1513.}
\label{tabapp}
\begin{tabular}{lcc}
\hline
Model & \texttt{relxilllp} & \texttt{relxillD} \\
\hline \noalign{\smallskip}
Galactic absorption (\texttt{tbabs}) & & \\
$N_{\rm H}$ (10$^{20}$\,cm$^{-2}$) & 3.7 & 3.7 \\[0.8ex]
Warm absorber (\texttt{xabs}) & & \\
$\log\xi_1$ & 2.0$\pm$0.1& 1.7$\pm$0.1\\
$N_{\rm H,1}$ (10$^{21}$\,cm$^{-2}$)& 5.8$\pm$1.1 & 4.3$\pm$1.1\\
$\log\xi_2$ & >\,2.9 & >\,2.5\\
$N_{\rm H,2}$ (10$^{21}$\,cm$^{-2}$)& 4.8$\pm$1.8 & 3.2$\pm$2.2\\[0.8ex]
\multicolumn{3}{l}{Direct\,+\,reflected continuum (\texttt{relxill$^*$})} \\
$\Gamma$&2.02$\pm$0.01&2.02$\pm$0.04\\
$h$ ($r_{\rm g}) $ &<\,2.1&--\\
$\mathcal{R}$&--&0.90$\pm$0.15\\
$q$  &--&>\,7.9\\
$a^*$ &>\,0.997&0.15$\pm$0.25\\
$\log\xi_{\rm d}$ &1.15$\pm$0.15&0.8$\pm$0.1\\
$\log\,(n/{\rm cm}^{-3})$&15\,(fixed)&>\,18.7\\
$A_{\rm Fe}$ &<\,0.62&3.3$\pm$1.1\\
$i$ (degrees)&43$\pm$2&38$\pm$3\\
$\log F_{\rm 0.3-10~keV}$&$-$11.04$\pm$0.01&$-$11.08$\pm$0.01 \\
\hline
C-stat/dof & 3245/3128 &3194/3126\\
\hline

\end{tabular}
\end{table}

\subsection{The 2015 \xmm observation}
\label{xmmobs}
\begin{figure}
   \centering
   \includegraphics[width=\columnwidth]{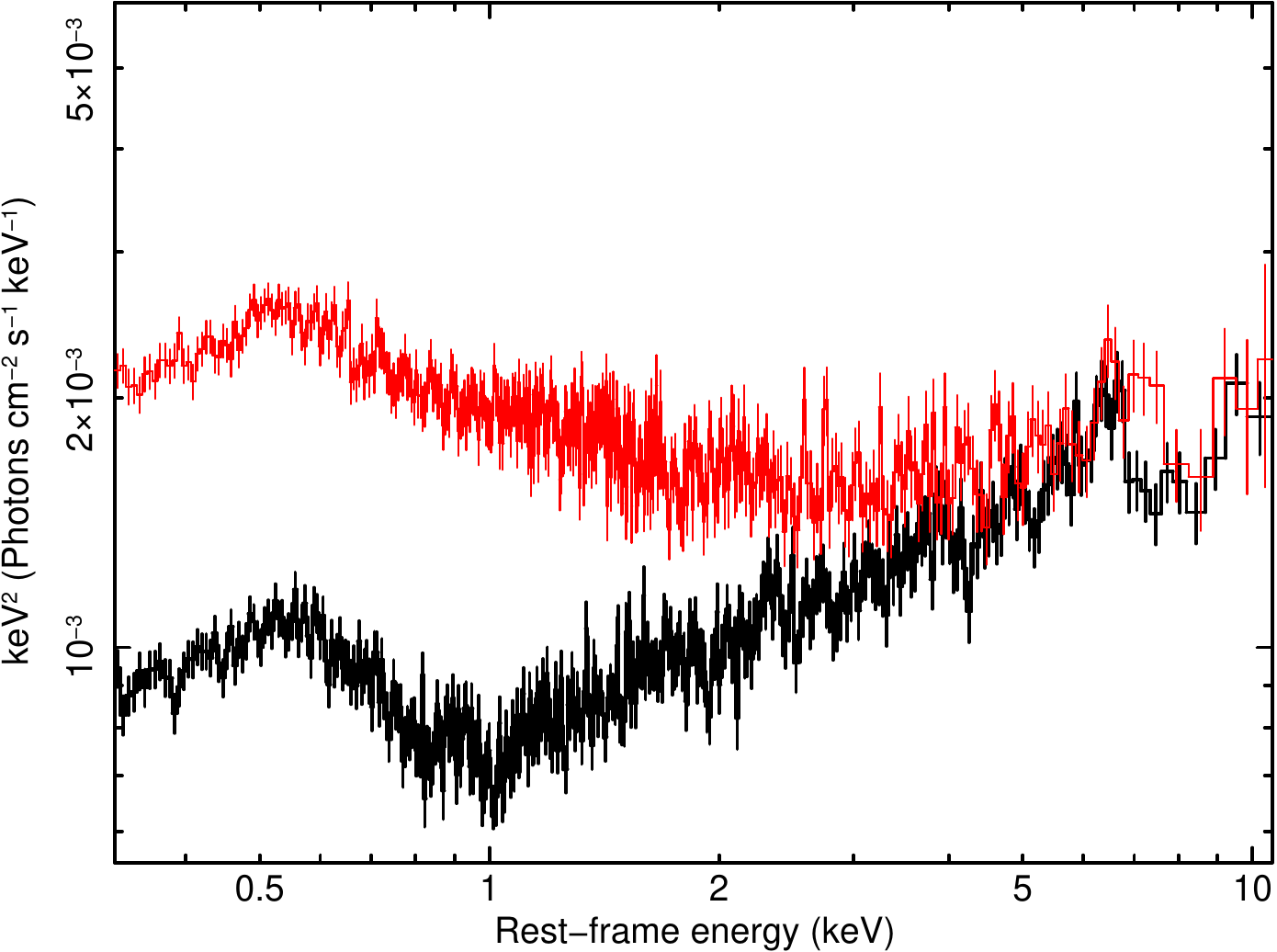}
      \caption{Unfolded EPIC spectra of Mrk\,1513 in the 2003 (black) and 2015 (red) observations. The soft X-ray spectrum shows a significant change, while the source is found in a similar spectral and flux state in the Fe-K band. The edge-like feature at $\sim$7--8 keV seems to be persistent, with an energy that possibly evolves with time.}
         \label{comparison}
   \end{figure}

\begin{figure}
\centering
\includegraphics[width=\columnwidth]{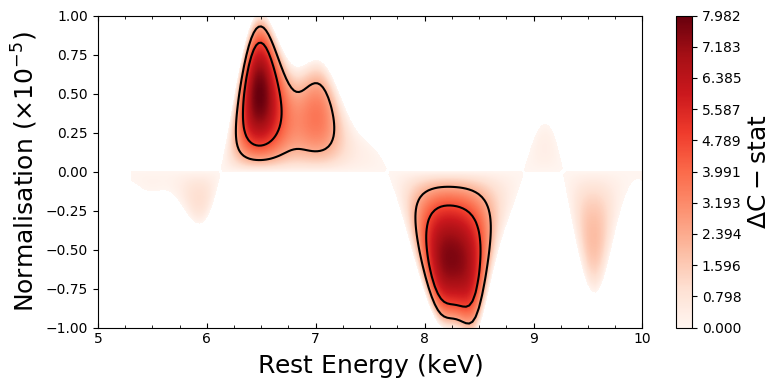}
 \caption{Same as Fig. \ref{linescan} for the 2015 \xmm observation of Mrk\,1513. The emission component and the secondary dip in the absorption profile appear to be rather stable in energy compared to 2003. The contours correspond to the 68\% and 90\% confidence levels, and the significance of both features is $\approx$2$\sigma$.}
\label{cont2}
\end{figure}

In the 2015 observation, the source lies in the gap between two chips of the EPIC-pn CCD. This translates into an extremely poor data quality above 5 keV, where the pn camera collected nearly four times less counts than in the 2003 exposure (see Table~\ref{tabobs}). In practice, the effective area of the pn in the Fe-K band in 2015 was equivalent to that a single MOS detector. Still, also this second \xmm observation provides some qualitative information about the possible Fe-K structure in this source. In Fig.~\ref{comparison} we show the 2003 and 2015 EPIC spectra obtained by combining the individual pn and MOS spectra using the \textsc{sas} routine \texttt{epicspeccombine}. The source is found in a strikingly different state, denoted by a much more prominent soft excess. On the other hand, the high-energy part of the 2015 spectrum shows milder changes with respect to the 2003 one. Remarkably, a deficit of net counts can be tentatively observed around $\sim$8 keV also in the second pointing of Mrk\,1513. There are again hints of a P-Cygni profile, although the emission and absorption component are suggested to be more detached by the blind line search, performed on the combined EPIC spectrum (Fig.~\ref{cont2}). Owing to the limited data quality, the statistical significance of both features is marginal, approaching the $\approx$95\% or 2$\sigma$ level. 

Despite the low statistics in the Fe-K band, we attempted to apply the \texttt{pcygx} and \texttt{diskline} models to the 2015 EPIC spectrum. The enhanced emission in the soft band was accounted for by adding a second phenomenological blackbody to the baseline continuum model. We refer to \citet{Jiang2022} for a possible description of the spectral variability by means of physically motivated models. The fits were performed with the same assumptions outlined in Sections \ref{ufo} and \ref{refl} obtaining, respectively, C-stat/d.o.f.\,=\,1190/1232 (\texttt{pcygx}) and 1186/1231 (\texttt{diskline}). The statistical improvement over the phenomenological continuum after the inclusion of the relevant Fe-K component is marginal in both cases, with $\Delta$C-stat/$\Delta$d.o.f.\,=\,$-$11/$-$3 in the disc-wind scenario and $-$15/$-$4 in the disc-line one. The key parameters of both components are loosely constrained yet generally consistent with the 2003 observation, as reported in Table~\ref{tabellaApp}.

\begin{table}
\caption{Best-fit parameters derived for the \texttt{pcygx} and \texttt{diskline} models adopted to reproduce the hard X-ray spectral feature(s) of Mrk\,1513 as observed in 2015. Errors are given at the 90\% confidence level for the single parameter of interest.} \label{tabellaApp}
\label{tabtext}
\begin{tabular}{lcc}
\hline
Model & P-Cygni & Disc line \\
\hline \noalign{\smallskip}
P-Cygni line (\texttt{pcygx}) & & \\
$E_{\rm 0}$ (keV) & 7.18$^{+0.31}_{-0.15}$ & -- \\
v$_{\inf}/c$ 
& 0.20$^{+0.25}_{-0.04}$ & -- \\
$\tau_{\rm tot}$ & 0.26$\pm$0.16 & -- \\[0.8ex]
Disc line (\texttt{diskline}) & & \\
$E_{\rm 0}$ (keV) & -- & 6.86$^{+0.15}_{-0.13}$ \\
$R_{\rm in}$ ($r_{\rm g}$) & -- & <\,275 \\
$i$ (degrees) & -- & >38 \\
$A$ (10$^{-5}$ cm$^{-2}$ s$^{-1}$) & -- & 1.2$\pm$0.6 \\
\hline
C-stat/d.o.f. & 1190/1232 & 1186/1231\\
\hline

\end{tabular}
\end{table}
%
\section{Discussion and Conclusions}

We have reported on the X-ray spectral properties of the highly accreting AGN Mrk\,1513, with specific focus on the peculiar edge-like feature discernible at $\sim$7 keV (Fig.~\ref{unfolded}). Despite the short duration (37.5 ks) of the 2003 \xmm observation, 
such a complex Fe-K emission and/or absorption structure is seen in the data from all the three EPIC detectors, and is tentatively present also in a lower quality spectrum that was serendipitously acquired 12 years later (Appendix \ref{xmmobs}). To date, these are the only observations of Mrk\,1513 carried out with the latest generation of X-ray facilities.

We have shown that, on sheer statistical grounds, it is not possible to discriminate between a wide-angle, ultra-fast wind and a relativistic accretion-disc line as the origin of the 7-keV feature. This notwithstanding, we can rely on some physical considerations to lean towards the most likely scenario. Indeed, in our analysis we have adopted a simplified phenomenological continuum, consisting of a soft X-ray blackbody and a hard X-ray power law, only modified by the known warm absorbers. However, the involved strength of any disc line, irrespective of its exact parameters, would suggest that X-ray reflection heavily contributes to the broadband spectral shape. When we switch to a physically motivated model for the disc-line interpretation, we find that the sharpness of the 7-keV feature is at odds with the overall spectral smoothness. This argument is not yet sufficient in itself to completely rule out the disc-line scenario, though, since a different physical component, such as warm Comptonisation, can partly or entirely account for the soft excess \citep[e.g.,][]{Middei2019,Matzeu2020}.

The nature of the broadband continuum is not as severe an issue in the disc-wind interpretation instead, as any emission/absorption wind signatures would be imprinted on the spectrum from the background X-ray source regardless of its shape and/or physical origin. In other words, the source of the continuum and that of the Fe-K feature(s) do not need to be one and the same. Hence, the P-Cygni solution remains fully viable at this stage, even in the absence of a self-consistent characterization of the entire spectrum. Interestingly, there is the hint of a change in the power-law photon index between the \texttt{pcygx} and \texttt{diskline} models, which is slightly higher (1.70 against 1.66; Table~\ref{tabella}) in the latter case. The steeper slope partly compensates for the lack of an absorption component, and suggests that the continuum trend above 10 keV could help discriminating between the two models. Unfortunately, with the current data quality, the number of significant spectral bins beyond the 7-keV edge is scarce, and it is not possible to reliably anchor the continuum level blueward of the apparent absorption trough. A similar situation was encountered after the first \xmm observation of the quasar PDS\,456, where a broad trough was detected extending across the upper end of the EPIC operating bandpass \citep{Reeves2003}. Only with the advent of \textit{NuSTAR}, which eventually probed the high-energy continuum, PDS\,456 has been conclusively established as the prototype of wide-angle disc winds in AGN \citep{Nardini2015}.
 
The disc-wind explanation is also appealing because of the unmistakable indications of a P-Cygni profile. Although the detection rate of blueshifted Fe-K features in the X-ray spectra of AGN is fairly high, leading to the notion that such outflows must cover a sizeable fraction of the solid angle, only a handful of P-Cygni profiles have been actually discovered so far. This, on the one hand, casts some doubts on the commonly accepted covering factor of ultra-fast X-ray winds among AGN and, on the other hand, prevents a better understanding of how the matter and energy flux from the central regions couples with the large-scale environment. At the time of the 2003 \xmm observation analysed here, a persistent outflow of highly ionised gas at $v_{\rm out}$\,$\sim$\,$-$1500 km s$^{-1}$ was present in Mrk\,1513, as revealed by the 1995, 1999, and 2010 {\it HST} observations \citep{Tofany2014}. The 2015 \xmm observation, despite its poor quality, provisionally suggests that also the X-ray wind might be persistent on similar timescales. Confirming the physical connection between the UV and X-ray outflows \citep[e.g.,][]{Mehdipour23} would shed new light on the relation between efficient accretion, disc winds, and large-scale feedback in AGN.
Our analysis has demonstrated that Mrk\,1513 is potentially a very promising source to investigate the AGN outflow phenomenon and, as such, it also stands out as a suitable target for the newly launched {\it XRISM} mission. 
Besides probing any possible fine structure in the Fe-K emission/absorption profiles, the high-resolution view would pinpoint the properties of the warm absorbers and reveal any residual continuum curvature, helping us to disentangle the contribution from disc (or even distant) reflection.
Longer X-ray exposures and, ideally, broader spectral coverage of this source are also desirable to achieve an accurate characterization of the disc wind properties and so determine its possible impact on the host galaxy. 




\begin{acknowledgements}
We thank the anonymous referee for their thorough reading of the manuscript. RM thanks Alessia Tortosa for useful suggestions and acknowledges financial support from the ASI--INAF agreement n. 2022-14-HH.0. This work relies on archival data, software or online services provided by the Space Science Data Center\,--\,ASI, and it is based on observations obtained with XMM-Newton, an ESA science mission with instruments and contributions directly funded by ESA Member States and NASA.

\end{acknowledgements}

%
%

%

\thispagestyle{empty}
\bibliographystyle{aa}
\bibliography{Mrk1315.bib}

\end{document}